\documentclass[10pt]{article}
\usepackage{epsfig}
\newcommand{\be}{\begin{equation}}
\newcommand{\ee}{\end{equation}}
\newcommand{\eq}[1]{(\ref{#1})}

\title{The influence of baryons
on the chiral phase transition in QCD.}
\author{M.N.~Chernodub and B.L.~Ioffe\\
\\
{\it\small Institute of Theoretical and Experimental Physics,}\\
{\it\small B.Cheremushkinskaya 25, 117218 Moscow,Russia}}

\date{}
\begin{document}
\maketitle
\def\la{\mathrel{\mathpalette\fun <}}
\def\ga{\mathrel{\mathpalette\fun >}}
\def\fun#1#2{\lower3.6pt\vbox{\baselineskip0pt\lineskip.9pt
\ialign{$\mathsurround=0pt#1\hfil##\hfil$\crcr#2\crcr\sim\crcr}}}
\begin{abstract}
A qualitative analysis of the chiral phase
transition in QCD with two massless quarks and non--zero baryon
density is performed. It is assumed that at zero baryonic density,
$\rho=0$, the temperature phase transition is of the second order.
Due to a specific power dependence of baryon masses on the chiral
condensate the phase transition becomes of the first order at the
temperature $T=T_{\mathrm{ph}}(\rho)$ for $\rho>0$. At
temperatures $T_{\mathrm{cont}}(\rho) > T > T_{\mathrm{ph}}(\rho)$
there is a mixed phase consisting of the quark phase (stable) and
the hadron phase (unstable). At the temperature $T =
T_{\mathrm{cont}}(\rho)$ the system experiences a continuous
transition to the pure chirally symmetric phase.
\end{abstract}
\bigskip


\bigskip

It is well known, that the chiral symmetry is valid in
perturbative quantum chromodynamics (QCD) with massless quarks. It
is expected also, that the chiral symmetry takes place in
full--perturbative and nonperturbative QCD at high temperatures,
($T \ga 200~\mbox{MeV}$), if heavy quarks ($c,b,t$) are ignored.
The chiral symmetry is strongly violated, however, in hadronic
matter, {\it i.e.} in QCD at $T=0$ and low density. What is the
order of phase transition between two phases of QCD with broken
and restored chiral symmetry at variation of temperature and
density is not completely clear now. There are different opinions
about this subject (for a detailed review see
Ref.~\cite{ref:1,ref:wil} and references therein).

We would like to consider here the  influence of baryon density on
the chiral phase transition in hadronic matter. Kogan, Kovner and
Tekin ~\cite{ref:a} have suggested the idea, that baryons may
initiate the restoration of chiral symmetry, if their density is
high -- when roughly half of the volume is occupied by baryons.
The physical argument in favour of this idea comes from the
hypothesis (supported by calculation in chiral soliton model of
nucleon~\cite{ref:b} ), that inside the baryon the chiral
condensate has the sign opposite to that in vacuum. This
hypothesis  is not proved. Even more, it is doubtful, that the
concept of quark condensate inside the nucleon can be formulated
in a correct way in quantum theory. But the idea on the strong
influence of baryon density on the chiral phase transition looks
very attractive. For this reason no assumption on the driving
mechanism of chiral phase transition at zero baryon density will
be done in this paper. The problem, under consideration is: how
the phase transition changes in the presence of baryons. The
content of the paper closely follows ref.5.

We discuss the phase transitions in QCD with two massless quarks,
$u$ and $d$. Many lattice calculations
\cite{ref:lattice:2nd:eta1,ref:lattice:2nd:noeta1,ref:lattice:2nd:noeta2,ref:lattice:2nd:eta2}
indicate, that at zero chemical potential the phase transition is
of the second order. It will be shown below, that the account of
baryon density drastically changes the situation and the
transition becomes of the first order, and, at high density, the
matter is always in the chirally symmetric phase.

Let us first consider a case of the zero baryonic density and
suppose that the phase transition from chirality violating phase
to the chirality conserving one is of the second order. The second
order phase transition is, generally, characterized by the order
parameter $\eta$. The order parameter is a thermal average of some
operator which may be chosen in various ways. The physical results
are independent on the choice of the order parameter. In QCD the
quark condensate, $\eta = \vert \langle 0 \vert \bar{u} u \vert 0
\rangle \vert = \vert \langle 0 \vert \bar{d} d \vert 0 \rangle
\vert \geq 0 $, may be taken as such parameter. In the confinement
phase the quark condensate is non--zero while in the deconfinement
phase it is vanishing.

The quark condensate has the desired properties: as it was
demonstrated in the chiral effective theory~\cite{ref:2,ref:3},
$\eta$ decreases with the temperature increasing and an extrapolation
of the curve $\eta(T)$ to the higher temperatures indicates,
that $\eta$ vanishes at $T=T^{(0)}_c \approx 180~\mbox{MeV}$. Here the superscript "0"
indicates that the critical temperature is taken at zero baryon density.
The same conclusion follows  from the lattice
calculations~\cite{ref:lattice:2nd:eta1,ref:lattice:2nd:eta2,ref:Karsch:Review},
where it was also found that the chiral condensate
$\eta$ decreases with increase of the chemical
potential~\cite{ref:mu1,ref:mu2}.

Apply the general theory of the second order phase
transitions~\cite{ref:6} and consider the thermodynamical
potential $\Phi(\eta)$ at the temperature $T$ near $T^{(0)}_c$.
Since $\eta$ is small in this domain, $\Phi(\eta)$ may be
expanded in $\eta$:
\be
\Phi(\eta) = \Phi_0 + \frac{1}{2} A \, \eta^2 + \frac{1}{4} B
\, \eta^4\,, \qquad B > 0\,.
\label{eq:1}
\ee
For a moment we neglect possible derivative terms in the potential.

The terms, proportional to $\eta$ and $\eta^3$ vanish for general
reasons~\cite{ref:6}. In QCD with massless quarks the absence of $\eta$
and $\eta^3$ terms can be proved for any perturbative Feynman
diagrams. At small $t = T - T^{(0)}_c$ the function $A(t)$ is linear in
$t$: $A(t) = a t$, $a>0$. If $t < 0$ the thermodynamical potential
$\Phi(\eta)$ is minimal at $\eta \neq 0$, while at $t > 0$ the
chiral condensate vanishes $\eta = 0$. At small $t$ the $t$-dependence of
the coefficient
$B(t)$ is inessential and may be neglected. The minimum, $\bar{\eta}$, of
the thermodynamical potential can be found from the condition,
$\partial \Phi/\partial \eta = 0$:
\be
\bar{\eta}=
\left\{
\begin{array}{ll}
\sqrt{-at/B}\,, \quad & t < 0\,; \\
0\,, \quad & t > 0\,.
\end{array}
\right.
\label{eq:2}
\ee
It corresponds to the second order phase transition since the potential is quartic in
$\eta$ and -- if the derivative terms are included in the expansion -- the correlation
length becomes infinite at $T=T^{(0)}_c$.

Turn now to the case of the finite, but small baryon density
$\rho$ (by $\rho$ we mean here the sum of baryon and
anti--baryon densities). For a moment, consider only one type
of baryons, {\it i.e.} the nucleon. The temperature of the
phase transition, $T_{\mathrm{ph}}$,
is, in general, dependent on the baryon density, $T_{\mathrm{ph}} =
T_{\mathrm{ph}}(\rho)$, with $ T_{\mathrm{ph}}(\rho=0) \equiv T^{(0)}_c$
At $T < T_{\mathrm{ph}}(\rho)$ the term, proportional to $E
\rho$, where $E=\sqrt{p^2 + m^2}$ is the baryon energy, must be
added to the thermodynamical potential~\eq{eq:1}. As was shown
in~\cite{ref:7,ref:8} the nucleon mass $m$ (as well as the
masses of other baryons) arises due to the spontaneous
violation of the chiral symmetry and is approximately
proportional to the cubic root of the quark condensate: $m = c
\eta^{1/3}$, with $c = (8 \pi^2)^{1/3}$ for a nucleon. At small
temperatures $T$ the baryon contribution to $\Phi$ is strongly
suppressed by the Boltzmann factor $e^{-E/T}$ and is
negligible. Below we assume that the proportionality $m \sim
\eta^{1/3}$ is valid in a broad temperature interval. Arguments
in favor of such an assumption are based on the expectation
that the baryon masses vanish at $T = T_{\mathrm{ph}}(\rho)$ and on the
dimensional grounds. Near the phase transition point $E =
\sqrt{p^2 + m^2} \approx p + c^2 \, \eta^{2/3} / (2 p)$. At
$\eta \to 0$ all baryons are accumulating near zero mass and a
summation over all baryons gives us -- instead of eq.~\eq{eq:1} --
the following:
\be
\Phi(\eta, \rho) = \Phi_0 + \frac{1}{2} a t \, \eta^2 +
\frac{1}{4} B \, \eta^4 + C \eta^{2/3} \rho\,,
\label{eq:3}
\ee
where $C = \sum_i c^2_i/(2 p_i)$. The term $\rho \sum_i p_i$
is absorbed into $\Phi_0$ since it is independent on the chiral
condensate $\eta$. The typical momenta are of the order of
the temperature, $p_i \sim T$. Thus, Eq.~\eq{eq:3} is valid in
the region $\eta \ll T^3$. In the leading approximation
the term $C$ can be considered as independent on the temperature at
$T \sim T^{(0)}_c$.

Due to the last term in Eq.~\eq{eq:3} the thermodynamical potential
{\it always} have a local minimum at $\eta=0$ since the condensate $\eta$
is always non--negative. At small $t<0$ there also exists a
local minimum at $\eta>0$, which is a solution of the equation:
\be
\frac{\partial \Phi}{\partial \eta} \equiv (a t + B \, \eta^2) \eta +
\frac{2}{3} C \rho \, \eta^{-1/3} = 0\,.
\label{eq:4}
\ee
\begin{figure}[!htb]
\begin{center}
\begin{tabular}{cc}
\includegraphics[angle=-00,scale=1.0,clip=true]{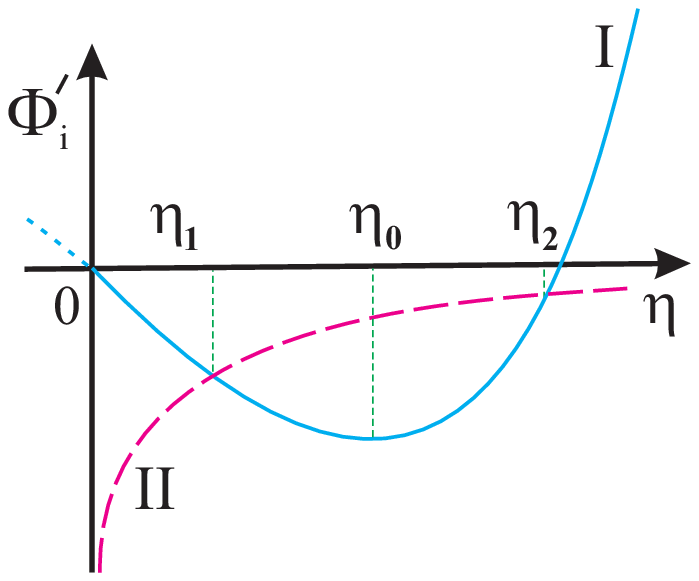} &
\includegraphics[angle=-00,scale=0.8,clip=true]{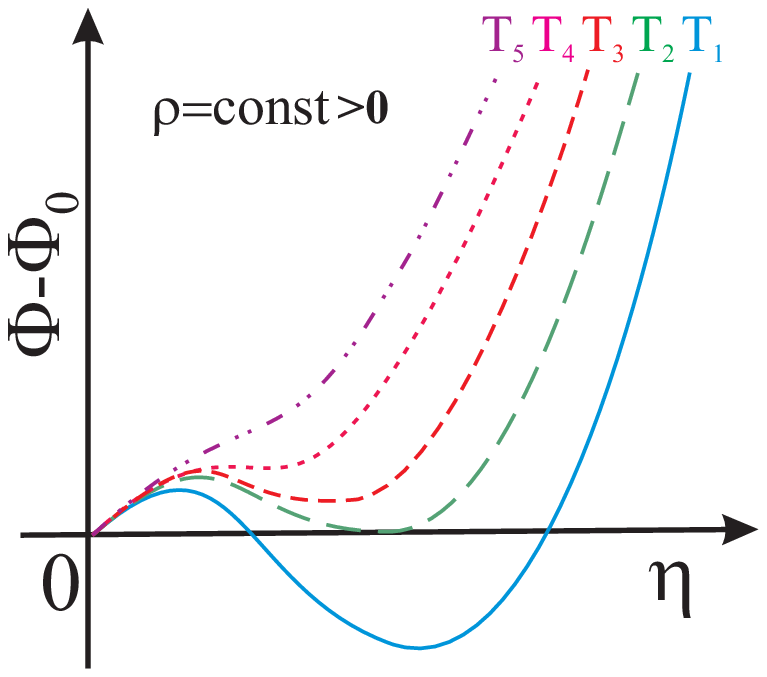} \vspace{1mm} \\
(a) & (b) \\
\end{tabular}
\end{center}
\caption{Figure 1: (a) Graphical representation of Eq.~\eq{eq:4}:
"I" is the first term and "II" the second term (with the
opposite sign) in the {\it r.h.s.} of the equation.
(b) The thermodynamic potential~\eq{eq:3} $vs.$
the chiral condensate at a fixed baryon density $\rho>0$.
At low enough temperatures, $T = T_1$, the system resides in the chirally
broken (hadron) phase. The first order phase transition to the quark phase
takes place at $T_{\mathrm{ph}} = T_2 > T_1$. At somewhat higher
temperatures, $T_3>T_{\mathrm{ph}}$ the system is in a mixed state.
The temperature $T_4 \equiv T_{\mathrm{cont}}$ corresponds to a continuous
transition to the pure quark phase, in which the thermodynamic potential has
the form $T_5$.}
\label{fig:roots}
\end{figure}
At small enough baryon density $\rho$, Eq.~\eq{eq:4} [visualized in Figure~\ref{fig:roots}(a)]
has, in general, two roots, $\eta_1<\eta_0$ and $\eta_2>\eta_0$, where
$\eta_0 = (- a t /3 B)^{1/2}$ is the minimum of the first term in
the right-hand side of Eq.~\eq{eq:4}.
The calculation of the second derivative
$\partial^2 \Phi/\partial \eta^2$ shows that the second root
$\eta_2$ (if exists) corresponds to minimum of $\Phi(\eta)$ and,
therefore, is a local minimum of $\Phi$. The point $\eta = \eta_1$ corresponds to
a local maximum of the thermodynamical potential since at this point
the second derivative is always non--positive.

The thermodynamical potential $\Phi(\eta, \rho)$ at (fixed) non--zero baryon
density $\rho$ has the form plotted in Figure~\ref{fig:roots}(b). At low
enough temperatures (curve $T_1$) the potential has a global minimum
at $\eta>0$ and system resides in the chirally broken (hadron) phase. As temperature
increases the minima at $\eta=0$ and at $\eta=\bar{\eta}_2>0$ becomes of equal
height (curve $T_2 \equiv T_{\mathrm{ph}}$). At this point the first order
phase transition to the quark phase takes place. At somewhat higher
temperatures, $T=T_3>T_{\mathrm{ph}}$, the $\eta>0$ minimum of the potential
still exist but $\Phi(\eta=0)<\Phi(\bar{\eta}_2)$. This is a mixed phase,
in which the bubbles of the hadron phase may still exist. However, as temperature
increases further, the second minimum disappears (curve
$T_4 \equiv T_{\mathrm{cont}}$). This temperature corresponds to a continuous
transition to the pure quark phase, in which the thermodynamic potential has
the form $T_5$.

Let us calculate the temperature of the phase transition, $T_{\mathrm{ph}}(\rho)$,
at non--zero
baryon density $\rho$. The transition corresponds to the curve $T_2$ in
Figure~\ref{fig:roots}(b), which is defined by the equation
$\Phi(\bar{\eta}_2,\rho)=\Phi(\eta=0,\rho)$, where $\bar{\eta}_2$
is the second root of Eq.~\eq{eq:4} as discussed above. The solution is
\be
T_{\mathrm{ph}}(\rho) = T_c^{(0)} -
\frac{5}{a} {\Biggl(\frac{2 \,C\, \rho}{3}\Biggr)}^{3/5}
\, {\Biggl(\frac{B}{4}\Biggr)}^{2/5}\,,
\label{eq:6}
\ee
and the second minimum of the thermodynamic potential is at
$\bar{\eta}_2 = {[4 a\,(T^{(0)} - T_{\mathrm{ph}}(\rho)) /(5\,B)]}^{1/2}$.

At a temperature slightly higher than $T_{\mathrm{ph}}(\rho)$ the potential
is minimal at $\eta = 0$, but it has also an unstable minimum
at some $\eta>0$. The existence of metastable state is also a
common feature of the first order phase transition ({\it e.g.},
the overheated liquid in case of liquid--gas system). With a
further increase of the density $\rho$ (at a given temperature)
the intersection of the two curves in Figure~\ref{fig:roots}(a)
disappears and the two curves only touch one another at one
point $\eta=\bar{\eta}_4$. At this temperature a continuous
transition (crossover) takes place. The corresponding potential
has the characteristic form denoted as $T_4$ in
Figure~\ref{fig:roots}(a). The temperature $T_4 \equiv
T_{\mathrm{cont}}$ is defined by the condition that the
first~\eq{eq:4} and the second derivatives of Eq.~\eq{eq:3}
vanish:
\be
T_{\mathrm{cont}}(\rho) = T_c^{(0)} - \frac{5}{a}\,
{\Biggl(\frac{2 \,C\, \rho}{9}\Biggr)}^{3/5}
\, {\Biggl(\frac{B}{2}\Biggr)}^{2/5}\,,
\label{eq:cross}
\ee
and the value of the chiral condensate, where the second local minimum of the
potential disappears is given by
$\bar{\eta}_4 = {[2 a (T_{\mathrm{cont}}(\rho) - T^{(0)}_c) /(5\, B)]}^{1/2}$.
At temperatures $T > T_{\mathrm{cont}}(\rho)$ the potential has only
one minimum and the matter is in the state with the restored chiral symmetry.
Thus, in QCD with massless quarks the type of phase transition with the restoration
of the chiral symmetry strongly depends on the value of baryonic density $\rho$.
At a fixed temperature, $T<T^{(0)}_c$, the phase transition happens at
a certain critical density, $\rho_{\mathrm{ph}}$. According to Eq.~\eq{eq:6}
the critical density has a kind of a "universal" dependence on the temperature,
$\rho_{\mathrm{ph}}(T) \propto  [T^{(0)}_c - T]^{5/3}$, the power of which does not
depend on the parameters of the thermodynamic potential, $a$ and $B$.

\begin{figure}[!htb]
\begin{center}
\begin{tabular}{cc}
\includegraphics[angle=-00,scale=1.0,clip=true]{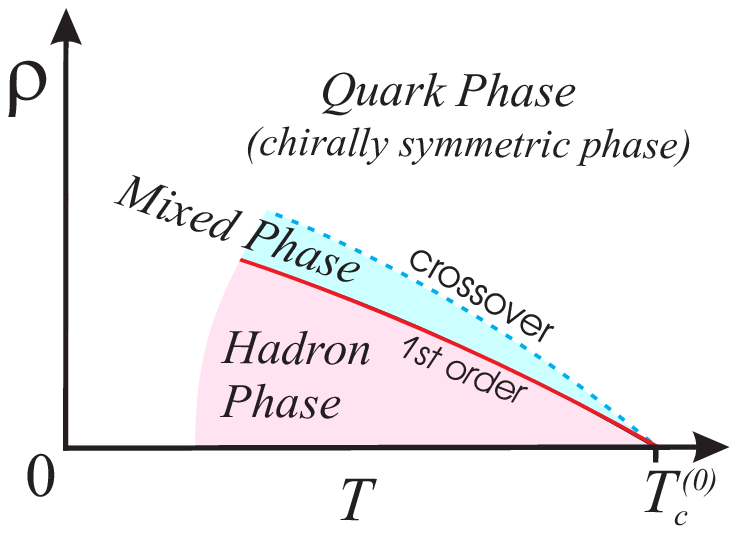} &
\includegraphics[angle=-00,scale=1.0,clip=true]{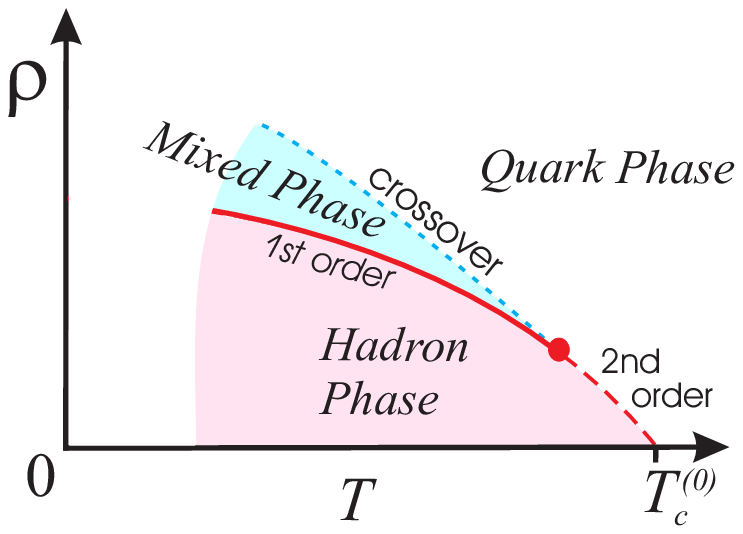} \\
(a) & (b)
\end{tabular}
\end{center}
\caption{Figure 2: The qualitative phase diagram at finite baryon density and
temperature based on the analysis (a) without and (b) with indication of the
approximate 2-nd order transition domain.}
\label{fig:phase}
\end{figure}

The expected phase diagram is shown qualitatively in Figure~\ref{fig:phase}(a).
This diagram does not contain an end-point which was found in lattice
simulations of the QCD with a finite chemical
potential~\cite{ref:endpoint1,ref:endpoint2}. We expect that
this happens because in our approach a possible influence of
the confinement on the order of the chiral restoration
transition was ignored. Intuitively, it seems that at low
baryon densities such influence is absent indeed: the
deconfinement phenomenon refers to the large quark--anti-quark
separations while the restoration of the of the chiral symmetry
appears due to fluctuations of the gluonic fields in the
vicinity of the quark. However, the confinement phenomenon
dictates the value of the baryon size which can not be ignored
at high baryon densities, when the baryons are overlapping. If
the melting of the baryons happens in the hadron phase depicted
in Figure~\ref{fig:phase}(a), then at high enough density the
nature of the transition could be changed.
This may give rise in appearance of the end-point observed in
Ref.~\cite{ref:endpoint1,ref:endpoint2}. The domain where the
inequality $|a t| \gg C \rho \eta^{2/3}$, $\rho \neq 0$ is fulfilled,
has specific features. In this domain the phase transition looks like a
smeared second order phase transition: the specific heat has
(approximately) a discontinuity at the phase transition point,
$\Delta C_p = a^2 T_c/B$. The correlation length increases as
${(T - T^{(0)}_c)}^{-1/2}$ at $T - T^{(0)}_c \to 0$. The latter arises if we
include the derivative terms in the effective thermodynamical potential.
The phase diagram with this domain indicated may look as it is shown
in Figure~\ref{fig:phase}(b).
Note that the applicability of our considerations is limited to the
region $|T - T_c^{(0)}|/T_c^{(0)} \ll 1$ and low baryon densities.

In the real QCD the massive heavy quarks (the quarks $c,b,t$)
do not influence on this conclusion, since their concentration
in the vicinity of $T \approx T_c^{(0)} \sim 200~\mbox{MeV}$ is
small. However, the strange quarks, the mass of which $m_s
\approx 150~\mbox{MeV}$ is just of order of expected $T_c^{(0)}$, may
change the situation. This problem deserves further investigation.

This work was supported in part by INTAS grant 2000-587, RFBR
grants 03-02-16209, 01-02-17456, 03-02-04016 and MK-4019.2004.2.

\end{document}